\documentclass{revtex4}
\usepackage{amssymb,amsfonts,amsmath,amsthm,mathtext,braket}

\begin{document}

\title{On constructing informationally complete covariant positive operator-valued measures
}

\author{\firstname{G.G.}~\surname{Amosov}}
\email[E-mail: ]{gramos@mi-ras.ru}

\address{Steklov Mathematical Institute of Russian Academy of Sciences, Moscow, Russia\\
Moscow Institute of Physics and Technology, Dolgoprudny, Russia\\
St. Petersburg State University, St.-Petersburg, Russia\\
Institute of Mathematics with Computer Center of the Ufa Science Center of Russian Academy of Sciences, Ufa, Russia}

\received{}

\begin{abstract} 
We study positive operator-valued measures generated by orbits of projective unitary representations of locally compact Abelian groups. It is shown that integration over such a measure defines a family of contractions being multiples of unitary operators from the representation. Using this fact it is proved that the measures are informationally complete. The obtained results are illustrated for the measure with density taking values in the set of coherent states. 
\end{abstract}

\keywords{projective unitary representation of locally compact Abelian group, positive operator-valued measures, informational completeness, the Naimark dilation, coherent states} 

\maketitle

\section{Introduction}

The theory of noncommutative operator-valued measures dating back to Naimark's pioneering work \cite{Naimark} has found important applications in quantum information theory. In particular, covariant measures defined by the symmetry group of a quantum system play an important role \cite{Hol}.
A natural way to construct covariant positive operator-valued measures (POVMs) is to integrate over orbits of an irreducible projective unitary representation of some group \cite{Hol, DAriano}. In the case of compact group, it is possible to obtain the measure possessing operator density with respect to the finite Haar measure on a group. If a group is locally compact, problems arise with the existence of an integral, which can be solved by using the Pontryagin duality principle \cite{Amo}. This topic is closely related to quantum tomography, which assumes not one, but a series of measurements determined by a set of different POVMs. In the context of exploiting the Pontryagin duality principle, such a method is proposed in \cite {Amo2}.

The paper is organized as follows. At first (Section 2) we provide the necessary information about the construction of POVM by means of orbits of the projective unitary representation introduced in \cite {Amo, Amo2}. In Section 3 we define a family of contractions associated with the constructed POVM and show that these contractions are multiples of unitary operators of the representation. Then, we prove that it results in the informationally completeness of POVM. Section 4 is devoted to an illustrative example in which a measure generated by coherent states is considered. The last Section contains concluding remarks.

\section{Preliminaries}

Let $G$ be a locally compact Abelian group with the Haar measure $\nu $. For the dual group $\hat G$ consisting of characters of $G$ we pick up the Haar measure $\hat \nu $ connected with $\nu $ by the Pontryagin duality
$$
\int \limits _{\hat G\times G}\chi (h)\overline {\chi (g)}\psi (g)d\hat \nu (\chi )d\nu (g)=\psi (h),\ \psi\in L^1(G,\nu ).
$$
Put $\mathfrak {H}=L^2(G,\mu )$ and define a set of unitary operators in $\mathfrak H$ by the formula
\begin{equation}\label{equ}
[U_{\chi ,g}\psi](h)=\chi (h)\psi(h+g),\ \chi \in \hat G,\ g,h\in G,\ \psi\in \mathcal {H}.
\end{equation}
The operators (\ref {equ}) are known to form a projective unitary representation of the group $\mathfrak {G}=\hat G\times G$ \cite{Amo}
\begin{equation}\label{com}
U_{\chi ,g}U_{\chi ',g'}=\chi '(g)U_{\chi \chi ',g+g'},\ \chi ,\chi '\in \hat G,\ g,g'\in G.
\end{equation}
It immediately follows from (\ref {com}) that
\begin{equation}\label{invers2}
U_{\chi ,g}U_{\chi ',g'}=\chi '(g)\overline {\chi (g')}U_{\chi ',g'}U_{\chi ,g}
\end{equation}
and
\begin{equation}\label{invers}
U_{\chi ,g}^*=\chi (g)U_{\overline{\chi },-g}.
\end{equation}

{\bf Proposition 1.} {\it The formula (\ref {equ}) determines an irreducible projective unitary representation of $\hat G\times G$ in $H=L^2(G,\mu)$.}

Proof.

Denote ${\mathfrak S}_2(H)$ the space of Hilbert-Schmidt operators in $H$.
It is known that the map $T:\ {\mathfrak S}_2(H)\to L^2(\hat G\times G)$ determined by the formula
$$
[T\rho ](\chi ,g)=Tr(\rho U_{\chi ,g}),\ \chi \in \hat G,\ g\in G,
$$
establishes an isometrical isomorphism \cite {Amo}. Moreover, the inverse transformation is given by
$$
T^{-1}F=\int \limits _{\hat G\times G}F(\chi ,g)U_{\chi ,g}^*d\hat \mu (\chi )d\mu (g),\ F\in L^2(\hat G\times G).
$$
Hence any finite dimensional projection $P$ belonging to $\mathfrak {S}_2(H)$ can be represented as
$$
P=\int \limits _{\hat G\times G}F_P(\chi ,g)U_{\chi ,g}^*d\hat \mu (\chi )d\mu (g)
$$
for some $F_P\in L^2(\hat G\times G)$. Thus, the claim $U_{\chi ,g}X=XU_{\chi ,g}$ for all $\chi \in \hat G,\ g\in G$ and a fixed bounded operator $X$ in $H$ results in $XP=PX$ for all finite dimensional projections $P$. The result follows.

$\Box $

Fix a unit vector $\psi _0\in \mathfrak {H}$.  Since the representation (\ref {equ}) is irreducible due to Proposition 1 the closer of $span(U_{\chi ,g}\psi _0,\ \chi \in \hat G,\ g\in G)$ coincides with $H$.
Denote $\Sigma$ and $B({\mathcal H})_+$ the $\sigma $-algebra of measurable subsets of $\mathfrak {G}$ and the positive cone of all positive bounded operators in $\mathcal H$.
Then, the map $\mathfrak {M}:\Sigma \to B({\mathcal H})_+$ determined by the formula
\begin{equation}\label{mera}
\mathfrak {M}(B)=\int \limits _B\ket {U_{\chi ,g}\psi _0}\bra {U_{\chi ,g}\psi _0}d\hat \nu (\chi )d\nu (g)
\end{equation}
is a covariant positive operator-valued measure \cite{Amo}.

\section{Informational completeness of $\mathfrak {M}$}

Following to the Naimark theorem \cite{Naimark} there exist the isometrical embedding ${\mathcal H}\subset {\mathcal K}$ and the projection-valued measure $E$ on $\mathfrak {G}$ such that
$$
\mathfrak {M}(B)=P_{\mathcal H}E(B)|_{\mathcal H},\ B\in \Sigma ,
$$
where $P_{\mathcal H}$ is the orthogonal projection on $\mathcal H$.
It implies that any unitary operator in $\mathcal K$ of the form
$$
W_{\chi ,g}=\int \limits _{\mathfrak G}\chi'(g)\overline {\chi (g')}dE(\chi ',g')
$$
determines the contraction $T_{\chi ,g}$ (i.e. an operator of norm at most one) in $\mathcal H$ by means of the formula
\begin{equation}\label{def}
T_{\chi ,g}=P_{\mathcal H}W_{\chi ,g}|_{\mathcal H}= \int \limits _{\mathfrak G}\chi'(g)\overline {\chi (g')}d\mathfrak {M}(\chi ',g')=\int \limits _{\mathfrak G}\chi'(g)\overline {\chi (g')}\ket {U_{\chi ',g'}\psi _0}\bra {U_{\chi ' ,g'}\psi _0}d\hat \nu (\chi ')d\nu (g')
\end{equation}
It is useful to remark that $W_{\chi ,g}$ is know as a unitary dilation of $T_{\chi ,g}$ and satisfies the relation \cite {NF} 
$$
T_{\chi ,g}^n=P_{\mathcal H}W_{\chi ,g}^n|_{\mathcal H},\ n=1,2,3,\dots
$$

{\bf Theorem 1.} {\it There exists a  complex-valued measurable functions $f$ on the group $\mathfrak {G}$ satisfying the relation $0<|f(\chi ,g)|\le 1,\ \chi \in \hat G,\ g\in G$, such that
$$
T_{\chi ,g}=f(\chi ,g)U_{\chi ,g},\ \chi \in \hat G,\ g\in G.
$$
}

{\bf Proof}.

Taking into account (\ref {com}), (\ref {invers}) and (\ref {def}) we obtain
$$
T_{\chi ,g}U_{\chi '',g''}=\int \limits _{\mathfrak G}\chi'(g)\overline {\chi (g')}\ket {U_{\chi ',g'}\psi _0}\bra {U_{\chi '',g''}^*U_{\chi ',g'}\psi _{0}}d\hat \nu (\chi ')d\nu (g')=
$$
$$
\int \limits _{\mathfrak G}\chi '(g)\overline {\chi (g')}\overline {\chi ''(g'')}\chi '(g'')\ket {U_{\chi ',g'}\psi _0}\bra {U_{\overline {\chi ''}\chi ',g'-g''}\psi _0}d\hat \nu (\chi ')d\nu (g')=
$$
$$
\int \limits _{\mathfrak G}\chi '\chi ''(g)\overline {\chi (g'+g'')}\overline {\chi ''(g'')}\chi '\chi ''(g'')\ket {U_{\chi '\chi '',g'+g''}\psi _0}\bra {U_{\chi ',g'}\psi _0}d\hat \nu (\chi ')d\nu (g')=
$$
$$
\chi ''(g)\overline {\chi (g'')}U_{\chi '',g''}\int \limits _{\mathfrak G}\chi '(g)\overline {\chi (g')}\ket {U_{\chi ',g'}\psi _0}\bra {U_{\chi ',g'}\psi _0}d\hat \nu (\chi ')d\nu (g')=
$$
$$
\chi ''(g)\overline {\chi (g')}U_{\chi '',g''}T_{\chi ,g},\ \chi ,\chi ''\in \hat G,\ g,g''\in G.
$$
Since the projective representation (\ref {equ}) is irreducible in $\mathcal H$ (\ref {invers2}) results in $T_{\chi ,g}=f(\chi ,g)U_{\chi ,g}$ by the Schur lemma.

$\Box $

The following statement shows that the measurement fulfilled by (\ref {mera}) is informationally complete.

{\bf Theorem 2.} {\it The density of probability distribution 
$$
p_{\rho}(\chi ,g)=\braket {U_{\chi ,g}\psi _0,\rho U_{\chi ,g}\psi _0}
$$
allows to restore a state $\rho $ by the formula
$$
\rho =\int \limits _{\mathfrak G}f^{-1}(\chi ,g)U_{\chi ,g}^*\int \limits _{\mathfrak G}\chi (g')\overline {\chi '}(g)p_{\rho }(\chi ',g')d\hat \nu (\chi ')d\nu (g')d\hat \nu (\chi )d\nu (g).
$$
}

{\bf Proof.}

At first, notice that
$$
\rho =\int \limits _{\mathfrak G}Tr(\rho U_{\chi ,g})U_{\chi ,g}^*d\hat \nu (\chi )d\nu (g)
$$
In fact, the map $\rho \to Tr(\rho U_{\chi ,g})$ determines the isometrical isomorphism \cite{Amo, Amo2}
$$
Tr(\rho \overline {\sigma })=\int \limits _{\mathfrak G}Tr(\rho U_{\chi ,g})\overline {Tr(\sigma U_{\chi ,g})}d\hat \nu (\chi )d\nu (g).
$$
It implies that
$$
\int \limits _{\mathfrak G}Tr(\rho U_{\chi ,g})Tr(U_{\chi ,g}^*\ket {\xi }\bra {\eta})d\hat \nu (\chi )d\nu (g)=
$$
$$
Tr(\rho \ket {\eta }\bra {\xi })=\braket {\xi ,\rho \eta}.
$$
Now it follows from Theorem 1 that
$$
Tr(\rho U_{\chi ,g})=f^{-1}(\chi ,g)\int \limits _{\mathfrak {G}}\chi '(g)\overline {\chi (g')}\braket {U_{\chi ',g'}\psi _0,\rho U_{\chi' ,g'}\psi _0}d\hat \nu (\chi ')d\nu (g').
$$

$\Box $

\section{The measure with density taking values in coherent states}

Consider $G={\mathbb R}$ giving $\hat G={\mathbb R}$ and ${\mathfrak G}=\hat G\times G={\mathbb R}^2$. Now ${\mathcal H}=L^2({\mathbb R})$ and
$$
(U_{x,y}\psi )(t)=e^{itx}\psi (t+y),\ (x,y)\in {\mathbb R}^2,\ \psi \in {\mathcal H}.
$$
Then,
$$
U_{x_1,y_1}U_{x_2,y_2}=e^{ix_2y_1}U_{x_1+x_2,y_1+y_2},\ x_j,y_j\in {\mathbb R},\ j=1,2,
$$
Put
$$
\psi _{0}(t)=\frac {1}{\pi ^{1/4}}exp(-t^2/2)
$$
{\bf Remark 1.} {\it 
Notice that $U_{x,y}\psi _0$ is a coherent state corresponding to the complex parameter $\alpha =\frac {-y+ix}{\sqrt 2}$ and after such a change of variables we obtain $D(\alpha )=e^{\frac {ixy}{2}}U_{x,y}$ (the displacement operator) \cite {Perelomov}. 
}

Taking into account
$$
\frac {1}{2\pi }\int \limits _{\mathbb R}e^{it(x-y)}dt=\delta (x-y)
$$
we obtain that the Haar measure corresponding to the Pontryagin duality \cite {Amo} is $\frac {1}{2\pi }dx$ and we are coming to the famous Glauber-Sudarshan measure
\begin{equation}\label{GS}
\mathfrak {M}(B)=\frac {1}{2\pi }\int \limits _B\ket {U_{x,y}\psi _0}\bra {U_{x,y}\psi _0}dxdy ,
\end{equation}
where $B$ runs measurable subsets of $\mathbb {R}^2$. Let us calculate (\ref {def}) appeared in Theorem 1 for (\ref {GS}).

{\bf Proposition 2.} {\it 
$$
T_{x,y}=exp\left (-\frac {x^2+y^2}{4}\right )e^{\frac {ixy}{2}}U_{x,y},\ x,y \in {\mathbb R}.
$$
}
{\bf Remark 2.} {\it Thus, the function $f(\chi ,g)\equiv f(x,y)$ defined in Theorem 1 is found for the case $G=\mathbb R$.}

{\bf Proof.}

Following to (\ref {def})
$$
T_{x,y}=\frac {1}{2\pi}\int \limits _{{\mathbb R}^2}e^{iry}e^{-ixt}\ket {U_{r,t}\psi _0}\bra {U_{r,t}\psi _0}drdt,
$$
Let us take a change of variables $\alpha =\frac {-y+ix}{\sqrt 2},\ \beta =\frac {-t+ir}{\sqrt 2}$, then 
$$
T_{\alpha }=\frac {1}{\pi}\int \limits _{{\mathbb R}^2}e^{2iIm(\alpha \overline {\beta})}\ket {\beta}\bra {\beta}d^2\beta .
$$

On the other hand, using the representation of the displacement operator in the form using the creation and annihilation operators $a^{\dag},\ a$ \cite{Perelomov}
$$
D(\alpha )=exp(|\alpha |^2/2)exp(-\overline {\alpha}a)exp(\alpha a^{\dag})
$$
we obtain
$$
\frac {1}{\pi }exp(|\alpha |^2/2)exp(-\overline {\alpha}a)\int \limits _{\mathbb C}\ket {\beta }\bra {\beta }d^2\beta exp(\alpha a^{\dag})=
$$
$$
exp(|\alpha |^2/2)\int \limits _{\mathbb C}exp(2iIm(\alpha \overline {\beta }))\ket {\beta }\bra {\beta }d^2\beta .
$$
Hence
$$
T_{\alpha }=exp(-|\alpha |^2/2)D(\alpha ).
$$

$\Box $

\section{Discussion}

In this paper we investigated a projective unitary representation of a direct product $\hat G\times G$, where $G$ is a locally compact Abelian group and $\hat G$ is its dual consisting of characters. Earlier it was shown \cite {Amo, Amo2} that orbits of this representation determines POVM using the Pontryagin duality. Continuing this study, we introduce a family of contractions associated with the constructed POVM. It is proved that these contractions are multiples of unitary operators of the representation. This fact allows us to prove the information completeness of the measure. The conversion formula restoring the state from the probability distribution on $\hat G\times G$ is obtained. In the last section, as an example, the case is analyzed when $G={\mathbb R}$ and the corresponding POVM is the Glauber-Sudarshan measure determined by projections on coherent states.

\end{document}